# Evidence for a Magnetic-Field induced Ideal Type-II Weyl State in Antiferromagnetic Topological Insulator Mn(Bi$_{1-x}$Sb$_x$)$_2$Te$_4$

Seng Huat Lee[1,2], David Graf[3*], Yanglin Zhu[1,2], Hemian Yi[2], Samuel Ciocys[4], Eun Sang Choi[3], Rabindra Basnet[5], Arash Fereidouni[5], Aaron Wegner[5], Yi-Fan Zhao[2], Lujin Min[2,6], Katrina Verlinde[6], Jingyang He[2,6], Ronald Redwing[1,6], V. Gopalan[6], Hugh O. H. Churchill[5], Alessandra Lanzara[4], Nitin Samarth[1,2], Cui-Zu Chang[2], Jin Hu[5*] and Z.Q. Mao[1,2*]

[1]2D Crystal Consortium, Materials Research Institute, The Pennsylvania State University, University Park, PA 16802, USA
[2]Department of Physics, The Pennsylvania State University, University Park, PA 16802, USA
[3]National High Magnetic Field Lab, Tallahassee, FL 32310, USA
[4]Department of Physics, University of California, Berkeley, Berkeley, CA 94720, USA
[5]Department of Physics, University of Arkansas, Fayetteville, AR 72701, USA
[6]Department of Materials Science and Engineering, The Pennsylvania State University, University Park, PA 16802, USA

The discovery of Weyl semimetals (WSMs) has fueled tremendous interest in condensed matter physics. WSMs require breaking of either inversion symmetry (IS) or time reversal symmetry (TRS); they can be categorized into type-I and type-II WSMs, characterized by un-tilted and strongly tilted Weyl cones respectively. Type-I WSMs with breaking of IS or TRS and type-II WSMs with IS breaking have been realized experimentally, but TRS-breaking type-II WSM still remains elusive. In this article, we report an ideal TRS-breaking type-II WSM with only one pair of Weyl nodes observed in the antiferromagnetic topological insulator Mn(Bi$_{1-x}$Sb$_x$)$_2$Te$_4$ under magnetic fields. This state is manifested by a large intrinsic anomalous Hall effect, a non-trivial $\pi$ Berry phase of **the cyclotron orbit and a large positive magnetoresistance in the ferromagnetic phase at an optimal sample composition. Our results establish a promising platform for exploring the physics underlying the long-sought, ideal TRS breaking type-II WSM.**

*Email: graf@magnet.fsu.edu; jinhu@uark.edu; zim1@psu.edu

# I. INTRODUCTION

Weyl semimetals (WSMs) have become a forefront research topic in contemporary condensed matter physics [1-3]. They provide model platforms for studying concepts in high-energy physics, such as the magnetic chiral anomaly effect. Additionally, they provide a means of realizing potentially useful exotic quantum states of technological relevance, *e.g.,* quantum anomalous Hall insulator [4]. In WSMs, singly-degenerate, linearly-dispersed bands cross at Weyl nodes in momentum space; these nodes always appear in pairs with opposite chirality and can be understood as monopole and anti-monopole of Berry flux. When the Weyl nodes are located at or near the chemical potential, their diverging Berry curvature can give rise to distinct properties such as the large intrinsic anomalous Hall effect (AHE) and anomalous Nernst effect if time-reversal symmetry (TRS) is broken [5-8]. Moreover, the bulk-edge correspondence principle leads WSMs to possess unique surface Fermi arcs [9-11], which can also generate exotic phenomena such as quantum oscillations [12,13] and a bulk quantum Hall effect through the formation of Weyl orbits [14]. Further understanding of Weyl fermion physics requires an *ideal* WSM in which all Weyl nodes should be symmetry related and at the same energy level (at or near the chemical potential), without interference from any other bands [1-3,15]. However, although ideal Weyl states have been long pursued, it has been realized only in bosonic systems (*i.e.*, photonic crystals) [16]. An ideal fermionic WSM with distinct exotic properties is still lacking.

The Weyl nodes of the WSMs discovered to date are either not close to the chemical potential, or not at the same energy level, or interfered with by other bands. These non-ideal WSMs can be divided into two categories: type-I and type-II WSMs. Type-I WSMs, which feature un-tilted Weyl cones, can be further categorized into inversion symmetry-breaking WSMs and TRS-breaking WSMs. The inversion symmetry-breaking type-I WSM state was first discovered in TaAs-class materials [10,11,17-19], and the TRS-breaking type-I WSM was recently demonstrated in several ferromagnetic (FM) materials such as $Co_3Sn_2S_2$ [5,6,20-22] and $Co_2MnGa$ [23,24]. Unlike type-I WSMs, type-II WSMs are characterized by strongly tilted Weyl cones and violate Lorentz invariance [25]. The inversion-symmetry-breaking type-II WSM was found in several non-magnetic materials such as $(W/Mo)Te_2$ [25-28], $LaAlGe$ [29], and $TaIrTe_4$ [30,31]; however, the TRS-breaking type-II WSM still remains elusive. $YbMnBi_2$ has been claimed to be a member of this category [32], but there has been debate regarding its TRS breaking [33]. In this article, we report experimental evidence for a recently-predicted, magnetic-field-induced *ideal* TRS breaking type-II Weyl state with only one pair of Weyl nodes in $MnBi_2Te_4$ [34,35].

$MnBi_2Te_4$ has been demonstrated as the first intrinsic antiferromagnetic (AFM) topological insulator (TI) [34-36]. The combination of spontaneous magnetization and non-trivial band topology makes this material accessible to a variety of exotic topological quantum states in its 2D thin layers, including the quantum anomalous Hall insulator [37], the axion insulator [38], and the $C = 2$ Chern insulator [39]. In addition, $MnBi_2Te_4$ is also predicted to host a long-sought, ideal type-II TRS breaking WSM when its AFM order is coerced to FM order under a magnetic field parallel to the *c*-axis [34,35]. Theoretical studies show that the interlayer hybridization, combined with the protection of $C_3$ rotational symmetry, induces the band crossings in the FM phase, resulting in a single pair of strongly tilted, type-II Weyl cones with the Weyl nodes at ~30 meV above the chemical potential [35]. Here, we provide experimental evidence for such an *ideal* WSM state using magnetotransport measurements in $Mn(Bi_{1-x}Sb_x)_2Te_4$. We achieve the ideal WSM by tuning the Sb:Bi ratio and observe the following key characteristics: an electronic phase transition across the AFM-to-FM phase boundary, a large intrinsic AHE, a non-trivial Berry phase of cyclotron orbit, and a large positive MR in the FM phase when the carrier density $n$ is close to a minimum. All these results provide strong support for the existence of the ideal Weyl state in this material.

## II. RESULTS AND DISCUSSION

Prior studies of Mn(Bi$_{1-x}$Sb$_x$)$_2$Te$_4$ [40,41] show that the carrier density is minimized near $x \sim 0.3$, where the carrier type also changes from electron to hole. Sb substitution for Bi also leads to striking changes in its magnetic properties through the Neel temperature shows a relatively small change from MnBi$_2$Te$_4$ ($T_N = 25$ K) to MnSb$_2$Te$_4$ ($T_N = 19$ K) [40,41]. Undoped MnBi$_2$Te$_4$ shows two magnetic transitions upon increasing the magnetic field along the *c*-axis, *i.e.*, the AFM to canted antiferromagnetic (CAFM) transition at $H_{c1}$ and the CAFM-to-FM transition at $H_{c2}$ [42]. Both $H_{c1}$ and $H_{c2}$ are suppressed by increasing Sb content and tend to merge as $x$ approaches 1[41]. Here, we focus on investigating the magnetotransport properties of Mn(Bi$_{1-x}$Sb$_x$)$_2$Te$_4$ under high magnetic fields (up to 41.5 T) and seek transport evidence for the predicted type-II WSM.

Single crystals of Mn(Bi$_{1-x}$Sb$_x$)$_2$Te$_4$ used in this work were grown by a flux method (see Methods). From magnetization measurements, we find the evolution of magnetism of Mn(Bi$_{1-x}$Sb$_x$)$_2$Te$_4$ with Sb content is consistent with prior studies [40,41]. Figure 1 shows the isothermal magnetization data at $T = 2$ K for several representative samples (blue curves). The AFM, CAFM, and FM phases determined by these data for each sample are color-coded in Fig. 1. We conducted high magnetic field Hall resistivity $\rho_{xy}$ measurements for these samples and find their magnetic field and temperature dependences of $\rho_{xy}$ exhibit drastic changes with increasing Sb content $x$. The slope of $\rho_{xy}$ shows negative-to-positive sign change near $x = 0.26$, indicating the carrier type changes to hole from electron for $x > 0.26$. The carrier type is sample dependent for the $x = 0.26$ batch. Although the $x = 0.26$(I-III) samples shown in Fig. 1 are from the same batch, the $x = 0.26$(I) sample is lightly electron-doped, while the $x = 0.26$(II-III) sample are lightly hole-doped.

To locate the position of the chemical potentials of the $x = 0.26$ samples, we performed the angle-resolved photoemission spectroscopy (ARPES) measurements (see Methods). We have measured three samples along the Γ-K direction selected from the $x = 0.26$ batch, denoted by $x = 0.26$(IV), 0.26(V), and 0.26(VI). The $x = 0.26$(IV) sample and the $x = 0.26$(I) sample used for transport measurements (Fig. 1b) were cut from the same piece. As shown in Fig. 2(a)(i), the chemical potential of the $x = 0.26$(IV) sample is very close to the bottom of the bulk conduction band, in a good agreement with the light electron doping probed in the transport measurements of the $x = 0.26$(I) sample as shown below. However, for the $x = 0.26$(V) and 0.26(VI) samples, we find their chemical potentials are pushed into the gap (Fig. 2(a)(ii-iii)). Note that the $x = 0.26$(IV-V) samples were measured at room temperature, while the measurements for the $x = 0.26$(VI) sample were carried out at 11 K, well below $T_N$ (=24.4 K). To compare with the lightly doped samples, we also measured the ARPES spectra along Γ-K of a moderately hole-doped sample with $x = 0.52$ (Fig. 2(a)(v)); its chemical potential is found to be well below the top of the bulk valence band. Additionally, our measurements also revealed features of the electronic inhomogeneity due to unavoidable chemical inhomogeneity. In the $x = 0.26$(VI) sample, we observed not only regions with chemical potential inside the gap [spot 1, Fig. 2(a)(iii)], but also regions with a clear Fermi edge close to the top of the bulk valence band [spot 2, Fig. 2(a)(iv)]. For spot 2, we have also constructed constant energy maps within two different energy windows (-300 meV to -260 meV and -40 meV to 0 meV) (Fig. 2(b)), from which we can see the projection of the Fermi surface on the $k_x$-$k_y$ plane is centered at Γ point. The transport properties of those lightly hole-doped samples shown in Fig. 1 ($x$ = 0.26(II-III)) should be dominated by regions of the sample in which the Fermi surface contains the point-like hole pocket (Fig. 2(a)(iv)), as supported by their magnetic field dependence of $\rho_{xy}$ (see Figs. 1(c-d) and Supplemental Material (SM) Section S1 [43]) and metal-like temperature dependences of resistivity $\rho_{xx}$ (SM Fig. S1).

From the field dependence of $\rho_{xy}$, we have estimated the carrier densities of those samples shown in Fig. 1. We note the heavily electron-doped (*e.g.* $x$ = 0) or hole-doped (*e.g.* $x$ = 0.79) samples exhibit linear magnetic field dependence of $\rho_{xy}$ in both the FM and PM phases (Figs. 1(a) and 1(f)). For the moderately doped samples ($x$ = 0.22 and 0.39), while they show quantum oscillations in $\rho_{xy}$ of the FM phases, their background field dependences remain linear (Figs. 1(a) and 1(e)). These observations are in stark contrast with the non-linear magnetic field dependence of $\rho_{xy}$ observed in lightly hole-doped samples ($x$ = 0.26(II-III), Figs. 1(c-d). The carrier density of those heavily/moderately doped samples can be estimated directly using the linear slope of $\rho_{xy}$ (*i.e.*, Hall coefficient $R_H$) and shows no or slight variation between the PM and FM states (Fig. 3). Note that the red triangles and purple diamonds in Fig. 3 represent the carrier densities of the samples in the PM states at 75 K and FM states at 2 K, respectively (see more $\rho_{xy}$ data of additional samples in Fig. S2 [43]). Fig. 3 shows the $x$ ~ 0.26 samples have a much lower carrier density; the minimum is ~3 × 10$^{18}$ cm$^{-3}$, about two orders of magnitude smaller than that of MnBi$_2$Te$_4$ [42,44]. For the $x$ = 0.26(II-III) samples which show the non-linear field dependence of $\rho_{xy}$ (Figs. 1(c-d)), their carrier densities are estimated through the two-band model fits (see SM Section S1 [43]). Furthermore, we found that the samples with lower carrier density have higher carrier transport mobility $\mu_H$. The maximal value of $\mu_H$ is ~1.1 × 10$^3$ cm$^2$/Vs (see Methods), about two orders of magnitude larger than those of MnBi$_2$Te$_4$ (~58 cm$^2$/Vs) and MnSb$_2$Te$_4$ (9.7 cm$^2$/Vs) (Fig. 3).

The access to the lower carrier density in $Mn(Bi_{1-x}Sb_x)_2Te_4$ enables the observation of the transport signatures of the predicted ideal FM WSM. As noted above, it has been theoretically predicted that the AFM-to-FM transition driven by a magnetic field along the out-of-plane direction should lead to a topological phase transition from an AFM TI to an FM WSM due to the crossing of spin-split bands induced by the FM exchange interaction [45]. Thus, an electronic phase transition (*i.e.*, band structure reconstruction) is expected to occur concomitantly with the AFM-CAFM-FM transition. This is indeed manifested by the magnetotransport properties of the lightly doped samples with $x = 0.26$(II-III). Unlike moderately/heavily doped samples ($x < 0.26$ or $> 0.26$) whose $R_H$ is hardly or weakly temperature-dependent below and above $T_N$ (Fig. 1(a), Figs. 1(e-f)), the lightly hole-doped samples ($x = 0.26$(II-III)) exhibit drastic temperature dependence in $\rho_{xy}(H)$ below 75 K (Figs. 1(c-d)). Moreover, with the magnetic field-driven AFM-CAFM-FM transition, the magnetic field dependences of $\rho_{xy}$ of these two samples show very unusual changes at $T = 2$ K, evolving from a nearly linear dependence in the AFM state to a small hump in the CAFM state and finally to a superlinear increase in the FM phase (see the red curves in Figs. 1(c-d)), in stark contrast with the sublinear increase in the PM state above $T = 75$ K. These observations imply the band structure of the FM state differs from that of PM, and the AFM-CAFM-FM transition drives a band structure reconstruction.

If this implicit electronic phase transition corresponds to the transformation from the AFM TI to the FM WSM as the theory predicts [34,35], typical transport signatures of an FM WSM such as a large intrinsic AHE and a $\pi$ Berry phase of cyclotron orbits should be probed through the transport measurements. In our experiments, we indeed observed these effects in the lightly and moderately doped samples. As indicated above, a large intrinsic AHE of an FM WSM arises from the diverging Berry curvature of the Weyl nodes, which is manifested by large intrinsic anomalous Hall conductivity $\sigma_{yx}^{AH}$ and anomalous Hall angle ($\Theta_{yx}^{AH} = \sigma_{yx}^{AH}/\sigma_{xx}$). Figure 4(a) presents $\sigma_{yx}^{AH}$ at $T = 2$ K for those representative samples shown in Fig. 1, which is derived from $\rho_{xy}$ and longitudinal resistivity $\rho_{xx}$ (SM Fig. S3 [43]) via tensor conversion (see Methods). We find $\sigma_{yx}^{AH}$ strongly depends on carrier type and density and shows values consistent with the theoretically predicted FM WSM state in moderately/lightly hole-doped samples. MnBi$_2$Te$_4$, which is heavily electron-doped, shows negative $\sigma_{yx}^{AH}$ and its absolute value $|\sigma_{yx}^{AH}|$ rises to a maximum of ~35 $\Omega^{-1}$cm$^{-1}$ within the CAFM phase, which can be attributed to the intrinsic AHE due to non-collinear spin structure [42]. However, in the FM phase, $|\sigma_{yx}^{AH}|$ drops to ~18 $\Omega^{-1}$cm$^{-1}$ and becomes nearly independent of magnetic field. For the lightly electron-doped sample ($x = 0.26$(I)), its $\sigma_{yx}^{AH}$ shows not only a sign reversal from being positive-to-negative in the CAFM phase, but also quantum oscillations in the FM phase, with its non-oscillatory component ~18 $\Omega^{-1}$cm$^{-1}$. In contrast, for heavily hole-doped samples, $\sigma_{yx}^{AH}$ shows very small positive values (*e.g.* $\sigma_{yx}^{AH}$ ~3 $\Omega^{-1}$cm$^{-1}$ for the $x = 0.79$ sample). However, when the hole carrier density is decreased to a moderate level, $\sigma_{yx}^{AH}$ of the FM phase increases significantly (*e.g.* the maximal $\sigma_{yx}^{AH}$ is ~42 $\Omega^{-1}$cm$^{-1}$ for the $x = 0.39$ sample) and also shows quantum oscillations. Remarkably, those lightly hole-doped samples ($x = 0.26$(II-III)) display very unusual field dependences: $\sigma_{yx}^{AH}$ strikingly varies with magnetic field in response to the AFM-CAFM-FM transition, with a positive-to-negative sign reversal occurring in the CAFM phase; its absolute value rises to a maximum at the CAFM-FM phase boundary (*e.g.* $|\sigma_{yx}^{AH}|_{max} \approx 66$ $\Omega^{-1}$cm$^{-1}$ for $x = 0.26$(II)). As indicated above, the ideal FM WSM state of MnBi$_2$Te$_4$ has only one pair of Weyl nodes near $\Gamma$, with the Weyl node separation $\Delta k$ being ~0.06 Å$^{-1}$ [34] (or 0.095 Å$^{-1}$ [35]) along the $k_z$ axis. $\sigma_{yx}^{AH}$ of such a Weyl state is proportional to $\Delta k$ (*i.e.* $\sigma_{yx}^{AH} = (e^2/2\pi h)\Delta k$) [46] and is estimated to 35 $\Omega^{-1}$cm$^{-1}$ (or 55 $\Omega^{-1}$cm$^{-1}$). The large $\sigma_{yx}^{AH}$ values seen in the FM phases of the moderately/lightly hole-doped samples are apparently consistent with such a predicted Weyl state.

To further examine the evolution of the AHE with the carrier density, we convert $\sigma_{yx}^{AH}$ to anomalous Hall angle $\Theta_{yx}^{AH}$ (see Methods and SM Fig. S4 [43]) and plot the maximal $\Theta_{yx}^{AH}$ or -$\Theta_{yx}^{AH}$ of the FM phase at $T = 2$ K as a function of the carrier density in Fig. 4(b), from which we can see $|\sigma_{yx}^{AH}|$ increases with reducing the carrier density for both electron and hole doping regimes. For the hole doping side, $|\sigma_{yx}^{AH}|$ increases to ~9 % for moderately hole-doped samples and up to a maximum of ~28 % for the lightly hole-doped sample of $x = 0.26$(II), comparable with those of recently-discovered magnetic WSMs such as Co$_3$Sn$_2$S$_2$ (~20 %) [6], Co$_2$MnGa (~12 %) [7] and GdPtBi (~16 %) [8]. These results imply that the Weyl nodes are near the chemical potential only when the Mn(Bi$_{1-x}$Sb$_x$)$_2$Te$_4$ system is tuned to be lightly/moderately hole-doped. The temperature dependences of $\rho_{xx}$ measured under magnetic fields for the lightly hole-doped samples also agree well for the expected field-induced FM WSM. As shown in Fig. 4(c), when the magnetic field is strong enough to drive the AFM state to the CAFM/FM state, $\rho_{xx}(T)$ exhibits an insulating-like behavior, in contrast with the metallic behavior under zero magnetic field. Given that the magnetic field-induced insulating-like behavior is commonly observed in topological semimetals due to their large magnetoresistance at low temperatures [15], our observation of such behavior is consistent with the predicted transition from the AFM TI to an FM WSM. As noted above, the FM WSM in MnBi$_2$Te$_4$ should belong to type-II [34,35] and has Weyl nodes at the touch points of electron and hole pockets. Therefore, the AFM-FM transition is expected to result in a Fermi surface transformation from one-hole pocket in the AFM state as revealed by ARPES measurements (Fig. 2(b)) to a Fermi surface consisting of both hole and electron pockets in the FM state. Such an expected Fermi surface reconstruction is evidenced by the positive-to-negative sign reversal of $\sigma_{yx}^{AH}$ in the CAFM phases of lightly doped samples. The striking magnetic field dependence of $\sigma_{yx}^{AH}$ in the FM phases of these samples suggests the Berry curvatures carried by the electron and the hole pockets have different magnetic field dependences. Further understanding of these results requires theoretical studies.

In addition to the large intrinsic AHE, the FM phases of the moderately/lightly doped samples are also found to show relativistic fermion behavior, manifested by non-trivial $\pi$ Berry phase of cyclotron orbits probed in Shubnikov–de Haas (SdH) oscillations. Figure 5(a) shows the transverse magnetoresistivity (MR) at various temperatures for the lightly electron-doped sample with $x = 0.26$(I), from which we can see the SdH oscillations appear following the CAFM-FM transition. Such SdH oscillations do not originate from topological surface states, but stem from the bulk bands of the FM state and can extend to the PM states below $T = 75$ K due to the PM's short-range magnetic order (see more discussions in SM Section S2 [43]). A single oscillation frequency of $F = 60$ T can be resolved by Fast Fourier transform (FFT) (Fig. 5(d)), thus Berry phases analyses can be conducted by constructing Landau level (LL) index fan diagram. Distinct from non-relativistic electron systems, relativistic fermions show a zeroth LL [47] associated with the Berry phase of $\pi$ [47], leading to a phase shift in quantum oscillations. As a result, an intercept of 0.5 is expected on the LL index $N$-axis by extrapolating the linear fit to the LL fan diagram. Note that the fan diagram is constructed here with integer LL indices being assigned to the minimal density of state at the Fermi level (DOS($E_F$)). Since SdH oscillations arise from oscillating scattering rate, conductivity should reach maxima at the minimal DOS($E_F$) where integer LL indices should be assigned [15] (see Methods). In Fig. 5(b), we present oscillatory conductivity $\sigma_{xx}^{osc}$ for the $x = 0.26$(I) and 0.39 samples derived via inverting the resistivity tensor and subtracting the non-oscillatory background. Following the above LL index definition, we plot the LL fan diagram in Fig. 5(c). The linear fits yield intercepts of $0.48 \pm 0.03$ and $0.64 \pm 0.06$ for the $x = 0.26$(I) and 0.39 samples respectively, which correspond to the non-trivial Berry phases of $(0.96 \pm 0.06)\pi$ and $(1.28 \pm 0.12)\pi$, indicating the carriers participating in the SdH oscillations on both samples are relativistic fermions. Furthermore, we also note the SdH oscillations show a systematic evolution with carrier density (see SM Fig. S5 [43]) and the oscillation frequency decreases with carrier density for both electron and hole doping sides (Fig. 5(d)). However, the lightly hole-doped samples showing a large AHE (see the red data points in Fig. 4(b)) do not show SdH oscillations, implying that its Weyl nodes are indeed near the chemical potential.

Another distinct magnetotransport property of a WSM is chiral anomaly, which should be manifested by negative longitudinal magnetoresistivity (LMR) [48,49]. However, in $Mn(Bi_{1-x}Sb_x)_2Te_4$, it is hard to distinguish the negative LMR due to the chiral anomaly from other sources of negative LMR, such as the suppression of spin scattering under magnetic fields in both CAFM and FM phases [42]. The other factor which prevents the clear observation of the chiral anomaly effect is that the lightly doped samples exhibit a large positive MR component, which likely overwhelms the negative LMR. In SM Fig. S6 [43], we present transverse and longitudinal magnetoresistance for both heavily and lightly doped samples. All heavily or moderately electron/hole-doped samples exhibit negative magnetoresistance arising from spin flip/flop transitions, with a maximum magnitude of ~25 % for $MnSb_2Te_4$ (SM Figs. S6(a) and 6(c)), suggesting that the transport in these samples is dominated by spin scattering. In contrast, the lightly doped samples exhibit large positive magnetoresistance (SM Figs. S6(b) and S6(d)). For instance, the sample showing the largest $\theta_{yx}^{AH}$ exhibits large positive MR, reaching ~180 % for $H \perp I$ and ~90 % for $H // I$ at 30 T and 2 K (SM Fig. S6(b) and S6(d)). All these observations indicate the scattering mechanism has changed essentially when the carrier density is tuned to be close to a minimum and are consistent with the expected large MR of a WSM.

We note the layered compound EuCd$_2$As$_2$ with an *A*-type AFM order has also recently been reported to host a WSM state with one pair of Weyl nodes either in its PM state [50] or field-driven FM phase [51,52]. As compared to EuCd$_2$As$_2$, the Weyl state in MnBi$_2$Te$_4$ has distinct natures: i) its Weyl cones are much more strongly tilted, forming a type-II WSM state; ii) the tunability of chemical potential by Sb substitution for Bi enables the observation of remarkable exotic quantum transport properties of Weyl fermions, including large intrinsic AHE and $\pi$ Berry phase.

## III. METHODS

### A. Sample preparation and measurements

Single crystals of Mn(Bi$_{1-x}$Sb$_x$)$_2$Te$_4$ with *x* from 0 to 1 were grown by a self-flux method as previously reported [44]. The mixtures of high purity manganese powder (99.95 %), antimony shot (99.9999 %), bismuth shot (99.999 %) and tellurium ingot (99.9999+ %) with the molar ratio of Mn:Sb:Bi:Te = 1:5$x$:5(1-$x$):16 were loaded into an alumina crucible and sealed in evacuated quartz tubes. The mixtures were heated up to 900 °C for 12 hours to promote homogeneous melting and slowly cooled down (1.5 °C/hour) to a temperature within the 590 °C – 630 °C range (depending on *x*), followed by centrifugation to remove excess flux. Higher temperatures were used for the samples with higher Sb content to avoid the formation of Sb$_2$Te$_3$ (melting point ~ 620 °C). Soft, plate-like single crystals up to $10 \times 5$ mm$^2$ can be obtained.

The phase purity of these single crystals is checked by X-ray diffraction. As shown in SM Fig. S7 [43], the sharp (00$L$) x-ray diffraction peaks, which show only small high angle shift from MnBi$_2$Te$_4$ to MnSb$_2$Te$_4$, demonstrate the excellent crystallinity and the formation of the desired crystal structure with the stacking of Te-(Bi,Sb)-Te-Mn-Te-(Bi,Sb)-Te septuple layers in our single crystal samples. The composition analyses by energy-dispersive X-ray spectroscopy (EDS) show the actual Sb content $x$ deviating from the nominal composition. In this paper, we used the measured Sb content $x$.

The low magnetic field magnetotransport and magnetization measurements were performed using a commercial Physical Property Measurement System (PPMS, Quantum Design). The high magnetic field transport and magnetization measurements were carried out using the 31 T, 35 T, 41.5 T resistive magnets at the NHMFL in Tallahassee. Standard four-probe samples are used for magnetotransport measurements, and the VSM method was used for high field magnetization measurements. The transport mobility $\mu_H$ is inferred from the Hall coefficient $R_H$ (= $\frac{d\rho_{xy}}{d(\mu_0 H)}$) via $\mu_H = R_H/\rho_{xx}(0)$ where $\rho_{xx}(0)$ is the zero-field longitudinal resistivity (shown in SM Fig. S2) [43]. The carrier density for the samples exhibiting linear field dependence in $\rho_{xy}$ is estimated via $n = \frac{1}{eR_H}$ where $e$ is the electric charge. For those samples showing non-linear $\rho_{xy}(H)$, two-band model fitting is used to estimate the carrier density and its mobility [43]. The 300 K ARPES measurements for the $x$ = 0.26(IV-V) and 0.52 samples were carried out using the 2DCC-MIP's ARPES facility at Penn State. The photoelectrons are excited by an unpolarized He-I$_\alpha$ light (~21.218 eV), and two analyzers (Scientia R3000 and DA30L) were used for these measurements. The $x$ = 0.26(VII) sample was measured at a LHe base temperature of 11 K with $p$-polarized 81 eV photons. These spectra were acquired at the Advanced Light Source Maestro

Beamline (BL 7.0.2) using the microARPES end station. The energy resolution is approximately 19 meV and the momentum resolution is 0.01 Å$^{-1}$.

### B. Analyses of anomalous Hall conductivity $\sigma_{yx}^{AH}$ and anomalous Hall angle $\theta_{yx}^{AH}$

We have extracted anomalous Hall conductivity $\sigma_{yx}^{AH}$ from measured Hall resistivity $\rho_{xy}$ and longitudinal resistivity $\rho_{xx}$ via resistivity-to-conductivity tensor conversion. For the samples showing linear field dependence in $\rho_{xy}$ in their FM states (*i.e.* $x$ = 0, 0.22, 0.26(I), 0.39 and 0.79, see Fig. 1), their $\sigma_{yx}^{AH}$ is derived using $\sigma_{yx}^{AH} = \rho_{xy}^{AH}/[\rho_{xx}^2+(\rho_{xy}^{AH})^2]$ where $\rho_{xy}^{AH}$ is anomalous Hall resistivity, obtained after subtracting the normal Hall contribution ($\rho_{xy}^{N} = R_H H$) from the measured $\rho_{xy}$. The $\rho_{xx}$ data of the samples shown in Fig. 1 are presented in SM Fig. S3 except for the $x$ = 0 sample whose $\rho_{xx}$ data were reported in our previous work [42]. For those lightly doped samples, since their $\rho_{xy}$ of FM states do not show linear field dependence, we extract $\sigma_{yx}^{AH}$ using $\sigma_{yx}^{AH} = \sigma_{yx} - \sigma_{yx}^{N}$ and $\sigma_{yx} = \rho_{xy}/(\rho_{xx}^2+\rho_{xy}^2)$. Here $\sigma_{yx}^{N}$ is the normal Hall conductivity, which can approximately be defined as the $\sigma_{yx}$ value at 24 K, slightly below $T_N$ (=24.4 K). In Fig. S8, we show the calculated $\sigma_{yx}$ of the $x$ = 0.26(II) sample as an example. The difference in $\sigma_{yx}$ between 2 K and 24 K represents $\sigma_{yx}^{AH}$. The anomalous Hall angle $\theta_{yx}^{AH}$ shown in Fig. 4(b) and SM Fig. S4 [43] is defined as $\sigma_{yx}^{AH}/\sigma_{xx}$ ; $\sigma_{xx}$ is also obtained via tensor conversion, *i.e.* $\sigma_{xx} = \rho_{xx}/(\rho_{xx}^2+\rho_{xy}^2)$.

### C. To assign LL indices for SdH oscillations

To obtain the correct Berry phase using the LL fan diagram, integer LL indices should be assigned when Fermi energy lies between LLs, *i.e.*, at the minimal DOS($E_F$) [53]. It should be emphasized that this general principle manifests differently between the quantum Hall effect (QHE) and SdH oscillations. For the QHE, minimal DOS($E_F$) occurs at the quantized Hall plateaus, where both longitudinal conductance and resistance reaches zero. The integer LL indices should be assigned to this state. However, as discussed in Ref. [15], the situation becomes different for SdH oscillations. Distinct from the *non-local* transport in a 2D quantum Hall system where the edge conduction channels are dominant, the resistance or conductance in conventional *diffusive* systems are governed by scattering. According to the transport theory, the scattering probability is proportional to the number of available states that electron can be scattered into [54,55]; thus, it oscillates in concert with the oscillations of DOS($E_F$) and gives rise to SdH oscillations [56,57], leading to conductivity maxima at the minimal DOS($E_F$), where the scattering rate is minimized. Therefore, the integer LL indices, which correspond to minimal DOS($E_F$), should be assigned to conductivity maxima in the SdH oscillations. This approach has been validated through comparative studies of the SdH and de Haas-van Alphen oscillations of topological semimetals such as ZrSiS [15,58].

**REFERENCES**

[1] B. Yan and C. Felser, *Topological Materials: Weyl Semimetals*, Annual Review of Condensed Matter Physics **8**, 337 (2017).

[2] N. P. Armitage, E. J. Mele and A. Vishwanath, *Weyl and Dirac semimetals in three-dimensional solids*, Reviews of Modern Physics **90**, 015001 (2018).

[3] A. Bernevig, H. Weng, Z. Fang and X. Dai, *Recent Progress in the Study of Topological Semimetals*, Journal of the Physical Society of Japan **87**, 041001 (2018).

[4] G. Xu, H. Weng, Z. Wang, X. Dai and Z. Fang, *Chern Semimetal and the Quantized Anomalous Hall Effect in $HgCr_2Se_4$*, Physical Review Letters **107**, 186806 (2011).

[5] Q. Wang *et al.*, *Large intrinsic anomalous Hall effect in half-metallic ferromagnet Co3Sn2S2 with magnetic Weyl fermions*, Nature Communications **9**, 3681 (2018).

[6] E. Liu *et al.*, *Giant anomalous Hall effect in a ferromagnetic kagome-lattice semimetal*, Nature Physics **14**, 1125 (2018).

[7] A. Sakai *et al.*, *Giant anomalous Nernst effect and quantum-critical scaling in a ferromagnetic semimetal*, Nature Physics **14**, 1119 (2018).

[8] T. Suzuki, R. Chisnell, A. Devarakonda, Y. T. Liu, W. Feng, D. Xiao, J. W. Lynn and J. G. Checkelsky, *Large anomalous Hall effect in a half-Heusler antiferromagnet*, Nat Phys **12**, 1119 (2016).

[9] X. Wan, A. M. Turner, A. Vishwanath and S. Y. Savrasov, *Topological semimetal and Fermi-arc surface states in the electronic structure of pyrochlore iridates*, Physical Review B **83**, 205101 (2011).

[10] S.-Y. Xu *et al.*, *Discovery of a Weyl Fermion semimetal and topological Fermi arcs*, Science **349**, 613 (2015).

[11] H. Weng, C. Fang, Z. Fang, B. A. Bernevig and X. Dai, *Weyl Semimetal Phase in Noncentrosymmetric Transition-Metal Monophosphides*, Physical Review X **5**, 011029 (2015).

[12] A. C. Potter, I. Kimchi and A. Vishwanath, *Quantum oscillations from surface Fermi arcs in Weyl and Dirac semimetals*, Nat Commun **5**, 5161 (2014).

[13] P. J. W. Moll, N. L. Nair, T. Helm, A. C. Potter, I. Kimchi, A. Vishwanath and J. G. Analytis, *Transport evidence for Fermi-arc-mediated chirality transfer in the Dirac semimetal $Cd_3As_2$*, Nature **535**, 266 (2016).

[14] C. Zhang *et al.*, *Quantum Hall effect based on Weyl orbits in $Cd_3As_2$*, Nature **565**, 331 (2019).

[15] J. Hu, S.-Y. Xu, N. Ni and Z. Mao, *Transport of Topological Semimetals*, Annual Review of Materials Research **49**, 207 (2019).

[16] B. Yang *et al.*, *Ideal Weyl points and helicoid surface states in artificial photonic crystal structures*, Science **359**, 1013 (2018).

[17] S.-M. Huang *et al.*, *A Weyl Fermion semimetal with surface Fermi arcs in the transition metal monopnictide TaAs class*, Nat Commun **6**, 7373 (2015).

[18] B. Q. Lv *et al.*, *Experimental Discovery of Weyl Semimetal TaAs*, Phys. Rev. X **5**, 031013 (2015).

[19] L. X. Yang *et al.*, *Weyl semimetal phase in the non-centrosymmetric compound TaAs*, Nat Phys **11**, 728 (2015).

[20] Q. Xu, E. Liu, W. Shi, L. Muechler, J. Gayles, C. Felser and Y. Sun, *Topological surface Fermi arcs in the magnetic Weyl semimetal $Co_3Sn_2S_2$*, Physical Review B **97**, 235416 (2018).

[21] D. F. Liu *et al.*, *Magnetic Weyl semimetal phase in a Kagomé crystal*, Science **365**, 1282 (2019).
[22] N. Morali *et al.*, *Fermi-arc diversity on surface terminations of the magnetic Weyl semimetal $Co_3Sn_2S_2$*, Science **365**, 1286 (2019).
[23] Z. Wang, M. G. Vergniory, S. Kushwaha, M. Hirschberger, E. V. Chulkov, A. Ernst, N. P. Ong, R. J. Cava and B. A. Bernevig, *Time-Reversal-Breaking Weyl Fermions in Magnetic Heusler Alloys*, Physical Review Letters **117**, 236401 (2016).
[24] I. Belopolski *et al.*, *Discovery of topological Weyl fermion lines and drumhead surface states in a room temperature magnet*, Science **365**, 1278 (2019).
[25] A. A. Soluyanov, D. Gresch, Z. Wang, Q. Wu, M. Troyer, X. Dai and B. A. Bernevig, *Type-II Weyl semimetals*, Nature **527**, 495 (2015).
[26] I. Belopolski *et al.*, *Discovery of a new type of topological Weyl fermion semimetal state in $Mo_xW_{1-x}Te_2$*, Nature Communications **7**, 13643 (2016).
[27] L. Huang *et al.*, *Spectroscopic evidence for a type II Weyl semimetallic state in $MoTe_2$*, Nat. Mater. **15**, 1155 (2016).
[28] K. Deng *et al.*, *Experimental observation of topological Fermi arcs in type-II Weyl semimetal $MoTe_2$*, Nat. Phys. **12**, 1105 (2016).
[29] S.-Y. Xu *et al.*, *Discovery of Lorentz-violating type II Weyl fermions in LaAlGe*, Science Advances **3**, e1603266 (2017).
[30] K. Koepernik, D. Kasinathan, D. V. Efremov, S. Khim, S. Borisenko, B. Büchner and J. van den Brink, *$TaIrTe_4$: A ternary type-II Weyl semimetal*, Physical Review B **93**, 201101 (2016).
[31] I. Belopolski *et al.*, *Signatures of a time-reversal symmetric Weyl semimetal with only four Weyl points*, Nature Communications **8**, 942 (2017).
[32] S. Borisenko *et al.*, *Time-Reversal Symmetry Breaking Type-II Weyl State in $YbMnBi_2$*, Nature Communications **10**, 3424 (2019).
[33] J.-R. Soh *et al.*, *Magnetic structure and excitations of the topological semimetal $YbMnBi_2$*, Physical Review B **100**, 144431 (2019).
[34] D. Zhang, M. Shi, T. Zhu, D. Xing, H. Zhang and J. Wang, *Topological Axion States in the Magnetic Insulator $MnBi_2Te_4$ with the Quantized Magnetoelectric Effect*, Physical Review Letters **122**, 206401 (2019).
[35] J. Li, Y. Li, S. Du, Z. Wang, B.-L. Gu, S.-C. Zhang, K. He, W. Duan and Y. Xu, *Intrinsic magnetic topological insulators in van der Waals layered $MnBi_2Te_4$-family materials*, Science Advances **5**, eaaw5685 (2019).
[36] M. M. Otrokov *et al.*, *Prediction and observation of an antiferromagnetic topological insulator*, Nature **576**, 416 (2019).
[37] Y. Deng, Y. Yu, M. Z. Shi, Z. Guo, Z. Xu, J. Wang, X. H. Chen and Y. Zhang, *Quantum anomalous Hall effect in intrinsic magnetic topological insulator $MnBi_2Te_4$*, Science, eaax8156 (2020).
[38] C. Liu *et al.*, *Robust axion insulator and Chern insulator phases in a two-dimensional antiferromagnetic topological insulator*, Nat Mater **19**, 522 (2020).
[39] J. Ge, Y. Liu, J. Li, H. Li, T. Luo, Y. Wu, Y. Xu and J. Wang, *High-Chern-Number and High-Temperature Quantum Hall Effect without Landau Levels*, National Science Review, nwaa089 (2020).
[40] B. Chen *et al.*, *Intrinsic magnetic topological insulator phases in the Sb doped $MnBi_2Te_4$ bulks and thin flakes*, Nature Communications **10**, 4469 (2019).

[41] J. Q. Yan, S. Okamoto, M. A. McGuire, A. F. May, R. J. McQueeney and B. C. Sales, *Evolution of structural, magnetic, and transport properties in MnBi$_{2-x}$Sb$_x$Te$_4$*, Physical Review B **100**, 104409 (2019).
[42] S. H. Lee *et al.*, *Spin scattering and noncollinear spin structure-induced intrinsic anomalous Hall effect in antiferromagnetic topological insulator MnBi$_2$Te$_4$*, Physical Review Research **1**, 012011 (2019).
[43] See Supplemental Material at [URL will be inserted by publisher] for details description.
[44] J. Q. Yan *et al.*, *Crystal growth and magnetic structure of MnBi$_2$Te$_4$*, Phys. Rev. Mater. **3**, 8, 064202 (2019).
[45] J. Li, C. Wang, Z. Zhang, B.-L. Gu, W. Duan and Y. Xu, *Magnetically controllable topological quantum phase transitions in the antiferromagnetic topological insulator MnBi$_2$Te$_4$*, Physical Review B **100**, 121103 (2019).
[46] A. A. Burkov, *Anomalous Hall Effect in Weyl Metals*, Physical Review Letters **113**, 187202 (2014).
[47] T. Ando, *Physics of Graphene: Zero-Mode Anomalies and Roles of Symmetry*, Progress of Theoretical Physics Supplement **176**, 203 (2008).
[48] Y.-S. Jho and K.-S. Kim, *Interplay between interaction and chiral anomaly: Anisotropy in the electrical resistivity of interacting Weyl metals*, Physical Review B **87**, 205133 (2013).
[49] D. T. Son and B. Z. Spivak, *Chiral anomaly and classical negative magnetoresistance of Weyl metals*, Physical Review B **88**, 104412 (2013).
[50] J.-Z. Ma *et al.*, *Spin fluctuation induced Weyl semimetal state in the paramagnetic phase of EuCd$_2$As$_2$*, Science Advances **5**, eaaw4718 (2019).
[51] L.-L. Wang, N. H. Jo, B. Kuthanazhi, Y. Wu, R. J. McQueeney, A. Kaminski and P. C. Canfield, *Single pair of Weyl fermions in the half-metallic semimetal EuCd$_2$As$_2$*, Physical Review B **99**, 245147 (2019).
[52] J. R. Soh *et al.*, *Ideal Weyl semimetal induced by magnetic exchange*, Physical Review B **100**, 201102 (2019).
[53] Y. Ando, *Topological Insulator Materials*, Journal of the Physical Society of Japan **82**, 102001 (2013).
[54] A. B. Pippard, *The Dynamics of Conduction Electrons* (Gordon and Breach, New York, 1965).
[55] A. B. Pippard, *Magnetoresistance in Metals* (Cambridge University Press, Cambridge, 1989).
[56] D. Shoenberg, *Magnetic Oscillations in Metals* (Cambridge Univ. Press, Cambridge, 1984).
[57] M. V. Kartsovnik, *High Magnetic Fields: A Tool for Studying Electronic Properties of Layered Organic Metals*, Chemical Reviews **104**, 5737 (2004).
[58] J. Hu, Z. Tang, J. Liu, Y. Zhu, J. Wei and Z. Mao, *Nearly massless Dirac fermions and strong Zeeman splitting in the nodal-line semimetal ZrSiS probed by de Haas-van Alphen quantum oscillations*, Physical Review B **96**, 045127 (2017).

**ACKNOWLEDGMENTS**

The study is based upon research conducted at The Pennsylvania State University Two-Dimensional Crystal Consortium–Materials Innovation Platform (2DCC-MIP), which is supported by NSF cooperative agreement DMR 1539916. Z.Q.M. acknowledges the support from the US National Science Foundation under grant DMR1707502. J.H. acknowledges the US Department of Energy (DOE), Office of Science, Basic Energy Sciences program under Award No. DE-SC0019467 for the support of magnetization, quantum oscillation, and 31 T high field transport measurements. The work at the National High Magnetic Field Laboratory is supported by the NSF Cooperative Agreement No. DMR 1644779 and the State of Florida. L.J.M., J.Y.H., and V.G.'s work is supported by the Penn State Center for Nanoscale Science, an NSF MRSEC under the grant number DMR-1420620. C.Z.C., H. Y, and Y. F. Z. acknowledge the support from the DOE grant (DE-SC0019064) and the Alfred P. Sloan Research Fellowship (the 300 K ARPES measurements). The low-temperature ARPES experiment of this work was supported by the U.S. Department of Energy, Office of Science, Basic Energy Sciences, Materials Sciences and Engineering Division under Contract No. DE-AC02-05-CH11231 within the Quantum Material Program (KC2202). This experiment was performed at the Advanced Light Source, a DOE Office of Science User Facility under contract no. DE-AC02-05CH11231.

**Figures Captions**

Fig. 1. Magnetic and transport properties of Mn(Bi$_{1-x}$Sb$_x$)$_2$Te$_4$. Magnetic field dependence of Hall resistivity $\rho_{xy}$ at various temperatures and magnetization $M$ at 2 K for Mn(Bi$_{1-x}$Sb$_x$)$_2$Te$_4$ with (a) $x = 0$ and 0.22 (b-d) 0.26 (e) 0.39 and (f) 0.79. The rose, blue, and orange regions refer to the AFM, CAFM, and FM phase regions determined by magnetization (blue curves). The heavily (a and f) and moderately (e) doped samples exhibit linear magnetic field dependence for $\rho_{xy}$ or linear $\rho_{xy}$ background, while the lightly doped $x = 0.26$(II and III) samples show non-linear $\rho_{xy}$ (c-d). The $x = 0.26$(I, II, and III) samples used in this study were from the same growth batch, but displays either electron-like (b) and hole-like (c and d) transport behavior in $\rho_{xy}(H)$ owing to inevitable chemical inhomogeneity as discussed in the text.

Fig. 2. Band structures of Mn(Bi$_{1-x}$Sb$_x$)$_2$Te$_4$ (a) ARPES spectra measured on the samples with (i) $x = 0.26$(IV) (lightly electron-doped), (ii-iii) $x = 0.26$(V-VI) (chemical potential inside the bulk bandgap), (iv) $x = 0.26$(VI) (lightly hole-doped), and (v) 0.52 (moderately hole-doped). The spectra in (i), (ii) and (v) were measured at 300 K, whereas the spectra in (iii) and (iv) were measured at 11 K on two different spots of an identical sample. All the data in panel (a) are taken along the Γ-K direction (see panel (b)). (b) ARPES constant energy maps at spot 2, acquired by integrating over -300 meV to -260 meV (left) and -40 meV to 0 meV (right). A point-like Fermi surface centered at Γ can be seen on the $k_x$-$k_y$ plane.

Fig. 3. Carrier density and transport mobility of Mn(Bi$_{1-x}$Sb$_x$)$_2$Te$_4$. Composition dependence of carrier density $n$ and transport mobility $\mu_H$ of Mn(Bi$_{1-x}$Sb$_x$)$_2$Te$_4$. For heavily- and moderately-

doped samples showing linear $\rho_{xy}$ (Fig. 1 and SM Fig. S2 [43]), $n$ is obtained from the linear slope of $\rho_{xy}(H)$ at $T = 2$ K for the FM phase (purple diamond) and 75 K for the PM phase (red triangle). The black and blue circles represent transport mobility $\mu_H$ ($=R_H/\rho_{xx}(0)$) for the FM and PM phases, respectively. For lightly hole-doped samples ($x = 0.26$) with non-linear $\rho_{xy}(H)$, $n$ (empty orange triangle) and $\mu_H$ (empty blue circle) are extracted from the two-band model fit (see SM Section 1 [43]), and only the data of dominant bands are presented here.

Fig. 4. Anomalous Hall effect in Mn(Bi$_{1-x}$Sb$_x$)$_2$Te$_4$. (a) Field dependence of anomalous Hall conductivity $\sigma_{yx}^{AH}$ for the samples with $x = 0$, 0.26(I), 0.26(II), 0.26(III), 0.39, and 0.79 at 2 K/0.7 K. For the $x = 0.26$(II) and 0.26(III) samples with minimum carrier densities, $\sigma_{yx}^{AH}$ is extracted using $\sigma_{yx}^{AH} = \sigma_{yx} - \sigma_{yx}^{N}$ and $\sigma_{yx} = \rho_{xy}/(\rho_{xx}^2 + \rho_{xy}^2)$, where $\sigma_{yx}^{N}$ is the normal Hall conductivity defined as the $\sigma_{yx}$ value at 24 K. For all other samples, $\sigma_{yx}^{AH}$ is derived using $\sigma_{yx}^{AH} = \rho_{xy}^{AH}/[\rho_{xx}^2 + (\rho_{xy}^{AH})^2]$ where $\rho_{xy}^{AH}$ is the anomalous Hall resistivity, obtained after subtracting the normal Hall contribution ($\rho_{xy}^{N} = R_H H$) (see Methods). (b) The maximal anomalous Hall angle $\Theta_{yx}^{AH} = \sigma_{yx}^{AH}/\sigma_{xx}$ in the FM phase as a function of carrier density, obtained from the field dependence of $\Theta_{yx}^{AH}$ shown in SM Fig. S4 [43]. All data points are obtained at 2 K, except for the $x = 0.39$ samples (0.7 K). The two different colors of the data points in this panel (purple and red data) refer to two different methods used to evaluate $\sigma_{yx}^{AH}$ as stated above. (c) Temperature dependence of longitudinal resistivity $\rho_{xx}$ for the $x = 0.26$(II) sample at various fields (applied along the out-of-plane direction).

Fig. 5. Quantum oscillations in Mn(Bi$_{1-x}$Sb$_x$)$_2$Te$_4$. (a) Transverse magnetoresistance of the $x = 0.26$(I) sample at various temperatures from $T = 2$ K to 120 K. SdH oscillations can be seen in the FM phase and extend to the PM state below 75 K. (b) Longitudinal conductivity oscillations of the $x = 0.26$(I) and 0.39 samples at 2 K/0.7 K. Integer LL indices are assigned to the maximal conductivity where DOS($E_F$) reaches the minimum (see Methods). (c) LL fan diagram built from conductivity oscillations shown in (b). The relatively low oscillation frequency allows for approaching very low LLs in both samples. The slopes of the fitted lines are $61.2 \pm 0.3$ T and $80.6 \pm 0.9$ T for $x = 0.26$(I) and 0.39 samples, respectively, agreeing well with the oscillation frequencies obtained by FFT. This result, together with the accessibility of very low LLs, guarantees the reliability of the Berry phase obtained from LL fan diagram [53]. (d) FFT of SdH oscillations for various moderately/lightly electron- and hole-doped samples.

**Figure 1**

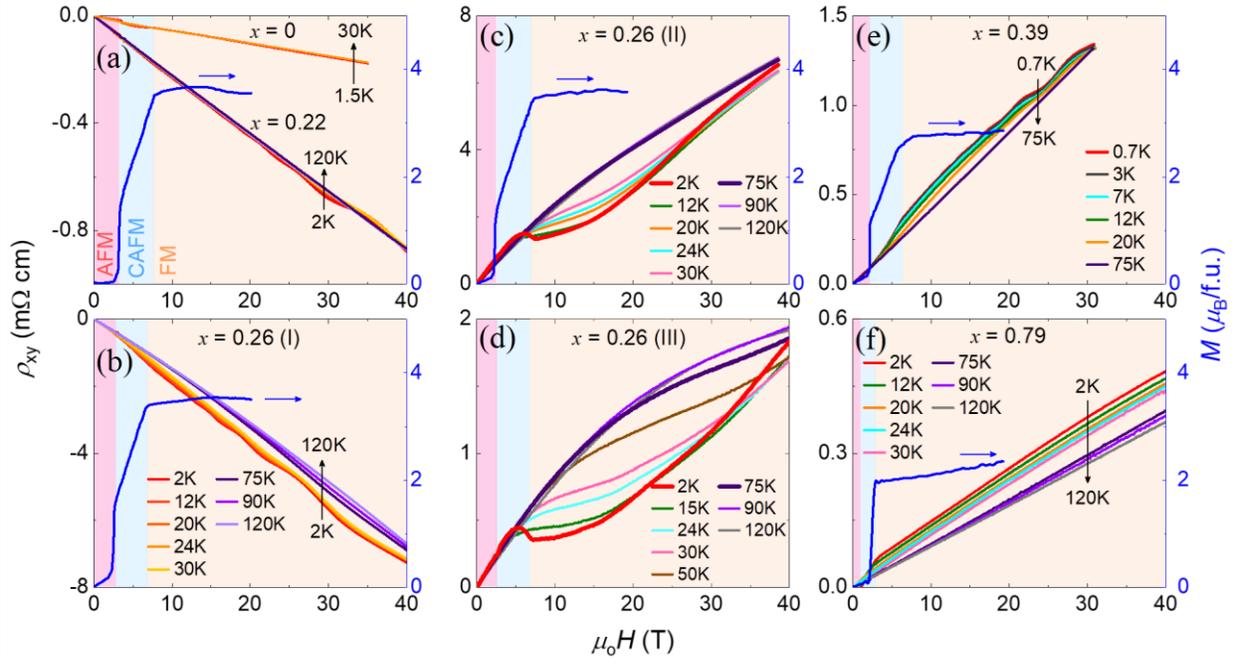

**Figure 2**

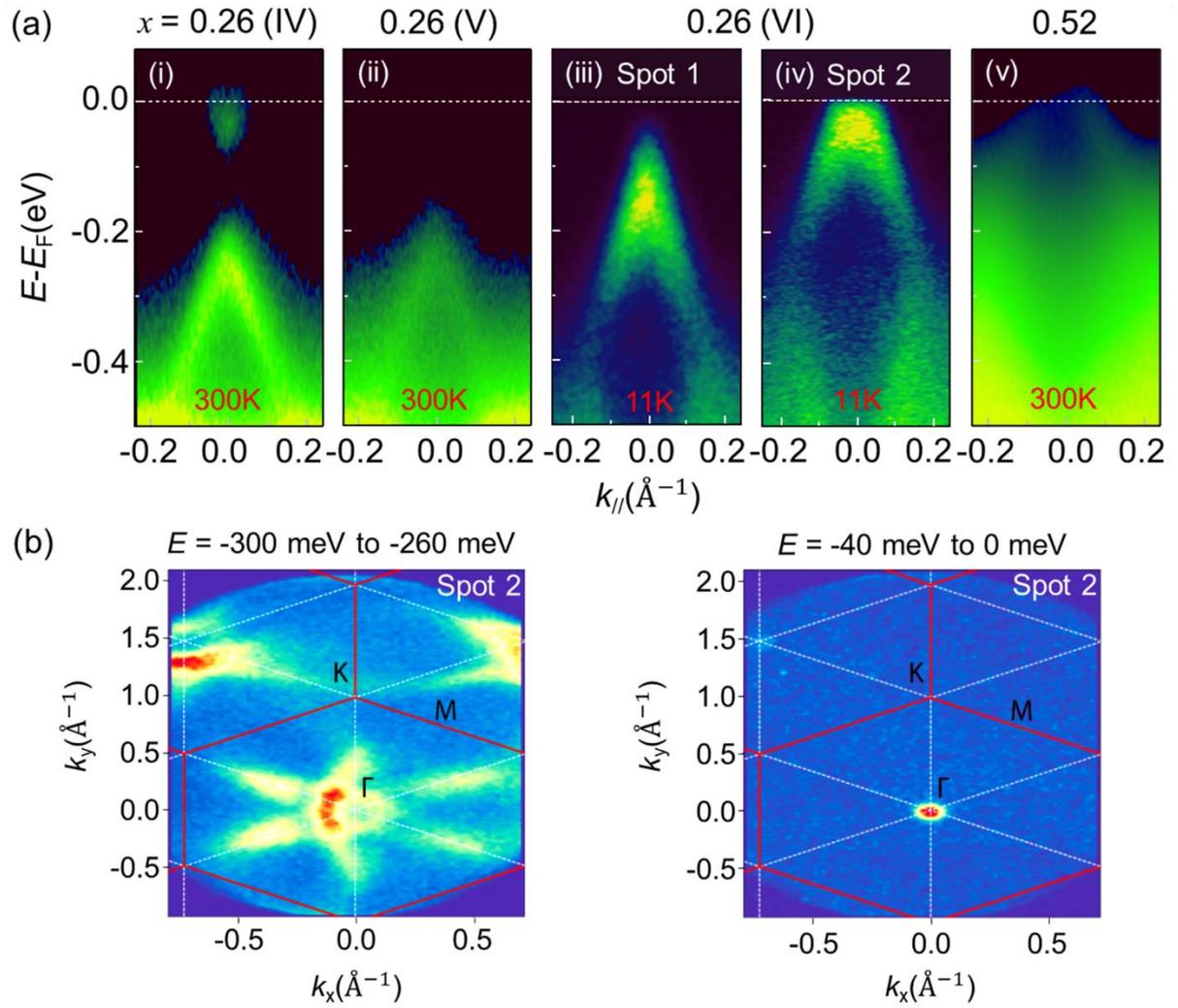

**Figure 3**

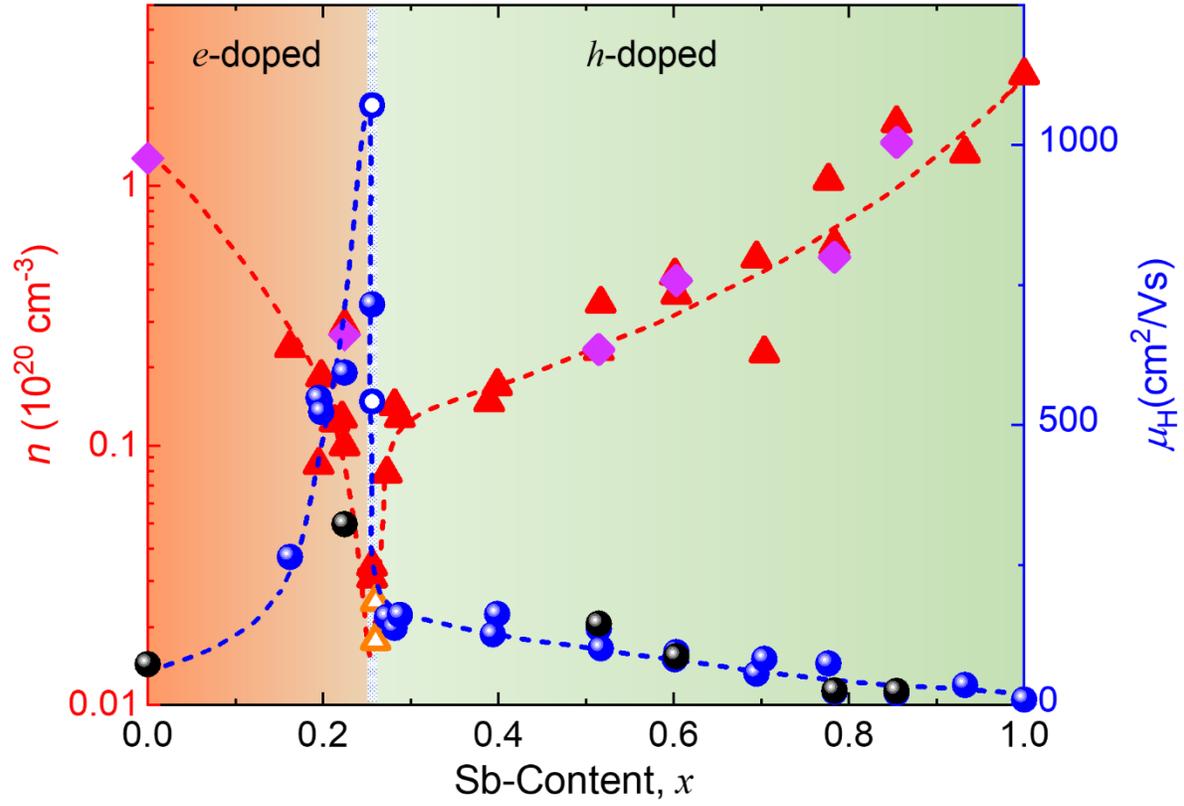

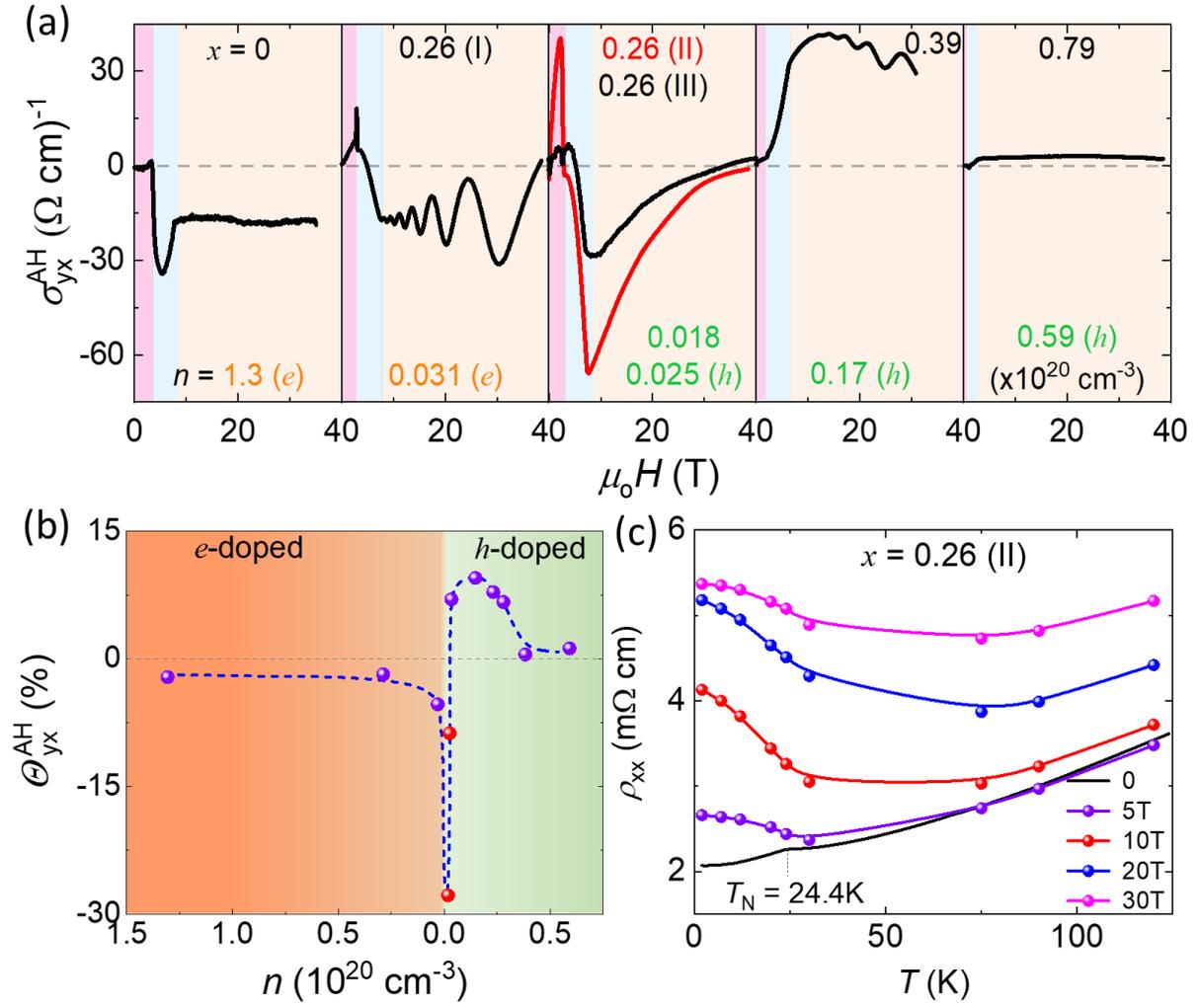

**Figure 5**

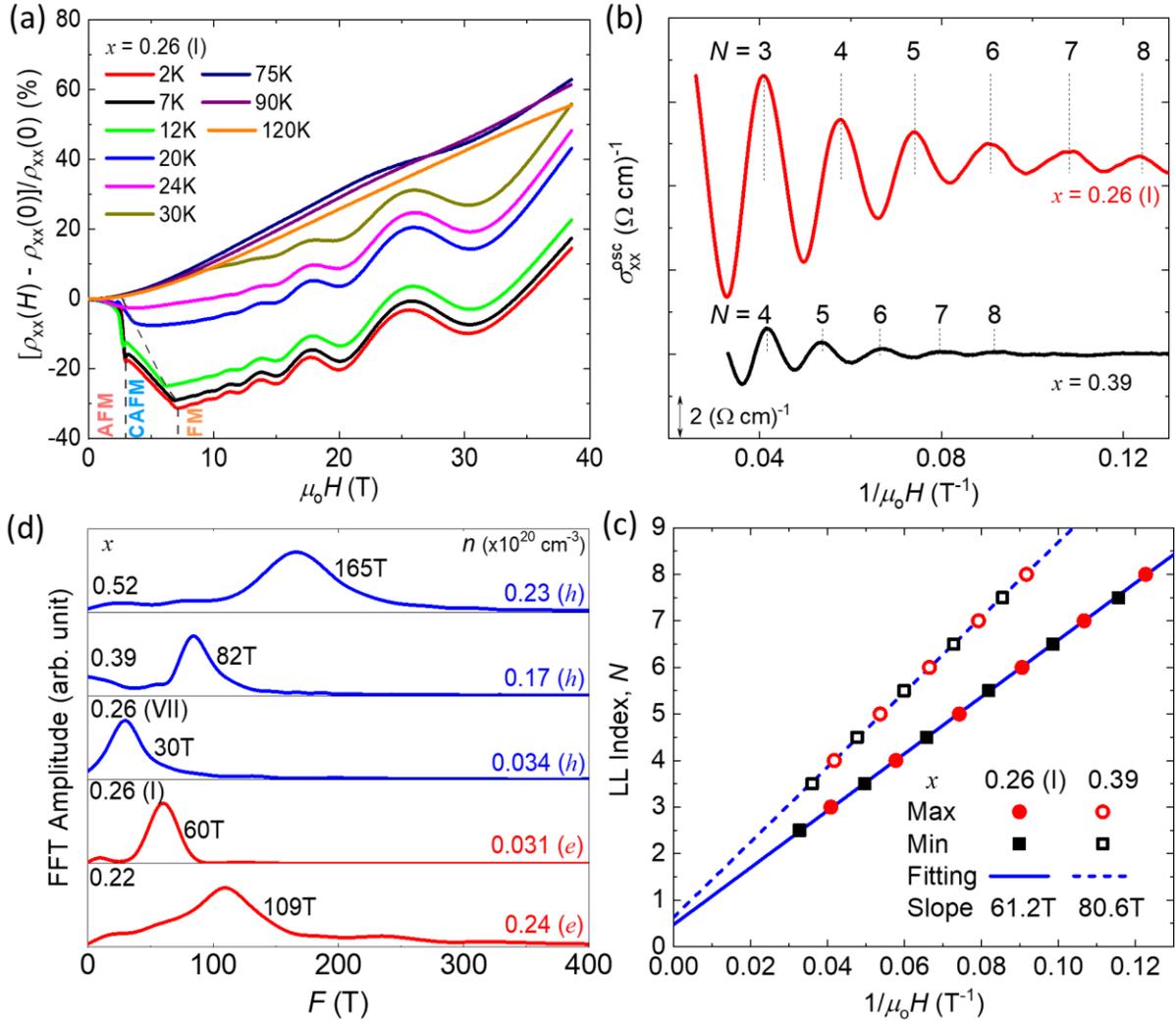